\documentstyle[prl,epsf,eqsecnum,aps,floats,amssymb,preprint]{revtex}
\tightenlines

\input epsf

\newcommand{\pbarp}{\mbox{$\bar{p}p$}}

\newcommand{\Et}{\mbox{$E_{T}$}}

\begin{document}
\preprint{FERMILAB-Pub-00/213-E, hep-ex/0008072}
\title{
The Ratio of Jet Cross Sections at
$\boldmath{\sqrt{s} = 630}$~GeV and 1800~GeV 
}
%
%
\author{                                                                      
B.~Abbott,$^{50}$                                                             
M.~Abolins,$^{47}$                                                            
V.~Abramov,$^{23}$                                                            
B.S.~Acharya,$^{15}$                                                          
D.L.~Adams,$^{57}$                                                            
M.~Adams,$^{34}$                                                              
G.A.~Alves,$^{2}$                                                             
N.~Amos,$^{46}$                                                               
E.W.~Anderson,$^{39}$                                                         
M.M.~Baarmand,$^{52}$
% ...
\\                                                                            
\vskip 0.30cm                                                                 
\centerline{(D\O\ Collaboration)}                                             
\vskip 0.30cm                                                                 
}                                                                             
\address{                                                                     
\centerline{$^{1}$Universidad de Buenos Aires, Buenos Aires, Argentina}       
\centerline{$^{2}$LAFEX, Centro Brasileiro de Pesquisas F{\'\i}sicas,         
                  Rio de Janeiro, Brazil}                                     
\centerline{$^{3}$Universidade do Estado do Rio de Janeiro,                   
                  Rio de Janeiro, Brazil}                                     
\centerline{$^{4}$Institute of High Energy Physics, Beijing,                  
                  People's Republic of China}                                 
\centerline{$^{5}$Universidad de los Andes, Bogot\'{a}, Colombia}             
\centerline{$^{6}$Charles University, Prague, Czech Republic}                 
\centerline{$^{7}$Institute of Physics, Academy of Sciences, Prague,          
                  Czech Republic}                                             
\centerline{$^{8}$Universidad San Francisco de Quito, Quito, Ecuador}         
\centerline{$^{9}$Institut des Sciences Nucl\'eaires, IN2P3-CNRS,             
                  Universite de Grenoble 1, Grenoble, France}                 
\centerline{$^{10}$CPPM, IN2P3-CNRS, Universit\'e de la M\'editerran\'ee,     
                  Marseille, France}                                          
\centerline{$^{11}$LPNHE, Universit\'es Paris VI and VII, IN2P3-CNRS,         
                  Paris, France}                                              
\centerline{$^{12}$DAPNIA/Service de Physique des Particules, CEA, Saclay,    
                  France}                                                     
\centerline{$^{13}$Panjab University, Chandigarh, India}                      
\centerline{$^{14}$Delhi University, Delhi, India}                            
\centerline{$^{15}$Tata Institute of Fundamental Research, Mumbai, India}     
\centerline{$^{16}$Seoul National University, Seoul, Korea}                   
\centerline{$^{17}$CINVESTAV, Mexico City, Mexico}                            
\centerline{$^{18}$FOM-Institute NIKHEF and University of                     
                  Amsterdam/NIKHEF, Amsterdam, The Netherlands}               
\centerline{$^{19}$University of Nijmegen/NIKHEF, Nijmegen, The               
                  Netherlands}                                                
\centerline{$^{20}$Institute of Nuclear Physics, Krak\'ow, Poland}            
\centerline{$^{21}$Institute for Theoretical and Experimental Physics,        
                   Moscow, Russia}                                            
\centerline{$^{22}$Moscow State University, Moscow, Russia}                   
\centerline{$^{23}$Institute for High Energy Physics, Protvino, Russia}       
\centerline{$^{24}$Lancaster University, Lancaster, United Kingdom}           
\centerline{$^{25}$University of Arizona, Tucson, Arizona 85721}              
\centerline{$^{26}$Lawrence Berkeley National Laboratory and University of    
                  California, Berkeley, California 94720}                     
\centerline{$^{27}$University of California, Davis, California 95616}         
\centerline{$^{28}$California State University, Fresno, California 93740}     
\centerline{$^{29}$University of California, Irvine, California 92697}        
\centerline{$^{30}$University of California, Riverside, California 92521}     
\centerline{$^{31}$Florida State University, Tallahassee, Florida 32306}      
\centerline{$^{32}$University of Hawaii, Honolulu, Hawaii 96822}              
\centerline{$^{33}$Fermi National Accelerator Laboratory, Batavia,            
                   Illinois 60510}                                            
\centerline{$^{34}$University of Illinois at Chicago, Chicago,                
                   Illinois 60607}                                            
\centerline{$^{35}$Northern Illinois University, DeKalb, Illinois 60115}      
\centerline{$^{36}$Northwestern University, Evanston, Illinois 60208}         
\centerline{$^{37}$Indiana University, Bloomington, Indiana 47405}            
\centerline{$^{38}$University of Notre Dame, Notre Dame, Indiana 46556}       
\centerline{$^{39}$Iowa State University, Ames, Iowa 50011}                   
\centerline{$^{40}$University of Kansas, Lawrence, Kansas 66045}              
\centerline{$^{41}$Kansas State University, Manhattan, Kansas 66506}          
\centerline{$^{42}$Louisiana Tech University, Ruston, Louisiana 71272}        
\centerline{$^{43}$University of Maryland, College Park, Maryland 20742}      
\centerline{$^{44}$Boston University, Boston, Massachusetts 02215}            
\centerline{$^{45}$Northeastern University, Boston, Massachusetts 02115}      
\centerline{$^{46}$University of Michigan, Ann Arbor, Michigan 48109}         
\centerline{$^{47}$Michigan State University, East Lansing, Michigan 48824}   
\centerline{$^{48}$University of Nebraska, Lincoln, Nebraska 68588}           
\centerline{$^{49}$Columbia University, New York, New York 10027}             
\centerline{$^{50}$New York University, New York, New York 10003}             
\centerline{$^{51}$University of Rochester, Rochester, New York 14627}        
\centerline{$^{52}$State University of New York, Stony Brook,                 
                   New York 11794}                                            
\centerline{$^{53}$Brookhaven National Laboratory, Upton, New York 11973}     
\centerline{$^{54}$Langston University, Langston, Oklahoma 73050}             
\centerline{$^{55}$University of Oklahoma, Norman, Oklahoma 73019}            
\centerline{$^{56}$Brown University, Providence, Rhode Island 02912}          
\centerline{$^{57}$University of Texas, Arlington, Texas 76019}               
\centerline{$^{58}$Texas A\&M University, College Station, Texas 77843}       
\centerline{$^{59}$Rice University, Houston, Texas 77005}                     
\centerline{$^{60}$University of Washington, Seattle, Washington 98195}       
}                                                                             

\date{\today}
\maketitle
\begin{abstract}
The D\O\ Collaboration has measured the inclusive jet cross section
in \pbarp\ collisions at $\sqrt{s} = 630$~GeV.  The results
for pseudorapidities $\left| \eta \right| <0.5$ are combined with 
our previous results at $\sqrt{s} = 1800$~GeV to 
form a ratio of cross sections with smaller uncertainties
than either individual measurement.  Next-to-leading-order QCD
predictions show excellent agreement with the measurement at 630~GeV; 
agreement is also satisfactory for the ratio.  Specifically,
despite a 10\% to 15\% difference in the absolute magnitude,
the dependence of the ratio on jet transverse momentum is very 
similar for data and theory.
\end{abstract}




For reactions with large momentum transfers, quantum chromodynamics (QCD)
treats complex proton-antiproton interactions in terms of simpler
scattering processes involving only one constituent from each particle.
Identifying these ``parton'' constituents with quarks and gluons, perturbative
QCD calculates production cross sections for scattered partons
(observed as showers or ``jets'' of collimated particles) that
also depend on empirically-determined parton distribution functions (PDF)
of the proton.

This measurement compares the production rate of 
jets as a function of their transverse energy, $E_{T}$, at two \pbarp\ 
center-of-mass energies: $\sqrt{s}=630$~GeV and 1800~GeV.  This
comparison reduces the systematic uncertainties and minimizes 
the prediction's sensitivity to choice of PDF.

In the simple parton model, inclusive jet cross sections scale 
with $\sqrt{s}$ in the sense that the dimensionless quantity 
$f(x_{T})=E_{T}^{4}\cdot E\frac{d^{3}\sigma }{d^{3}p}$, as a function of 
jet $x_{T}\equiv \frac{2E_{T}%
}{\sqrt{s}}$, does not depend on $\sqrt{s}$ \cite{feynman}.  
In this model, the ratio of scaled cross sections for different energies
is unity for all $x_{T}$. Although previous data
\cite{UA2_ratio,CDF_ratio} exhibited significant deviation 
from this naive scaling, the dimensionless framework 
provides a useful context for comparison with QCD. 
The D\O\ collaboration at the Fermilab Tevatron recently published
the inclusive jet cross section at $\sqrt{s}=1800$~GeV using $95700\pm 500$~nb$^{-1}$ 
of data \cite{xs_1800}. This Letter presents our complementary 
measurement at $\sqrt{s}=630$~GeV, using a sample of  $538\pm 22$~nb$^{-1}$
of data \cite{lum630}.  Because the data at both values of $\sqrt{s}$ were collected with
the same detector \cite{detector}, many uncertainties in the results are
highly correlated, and the ratio of the cross sections has greater precision
than either of the absolute measurements. 

The differential jet cross section, $d^{2}\sigma / dE_{T} d\eta$, is 
measured in bins of $E_{T}$ and pseudorapidity, 
$\eta \equiv -\ln \left(\tan \frac{\theta }{2}\right)$, where $\theta$
is the polar angle of the jet relative to the proton beam.  (In this
formulation, the dimensionless cross section, averaged over azimuth, 
is $\frac{E_{T}^{3}}{2\pi} \cdot d^{2}\sigma / dE_{T} d\eta$.)
The D\O\ reconstruction algorithm defines a jet by the total $E_{T}$ 
observed in calorimeter cells contained within
a cone of radius ${\cal R} \equiv [\left( \Delta \eta \right) ^{2}+\left( \Delta \phi 
\right) ^{2}]^{\frac{1}{2}}=0.7$, where $\phi$ is the azimuthal angle.   
When two such clusters of cells overlap, they are merged into a single jet if 
they share more than 50\% of the $E_{T}$ of the lower-$E_{T}$ cluster; 
otherwise, they are split into two separate jets, each defined by
its own $\eta$-$\phi$ centroid and $E_{T}$ value \cite{reco}.

The online trigger requires at least one jet above a set threshold.
The offline data selection procedure, which
suppresses backgrounds from electrons, photons, noise, and cosmic rays,
follows the methods used in the 1800~GeV analysis \cite{daniel,krane}. 
The efficiency of jet selection is approximately $96\%$ and is nearly independent
of jet $E_{T}$. To maintain precision in jet $E_{T}$, 
a vertex requirement removes jets resulting from
\pbarp\ interactions more than 50 cm from the center of the
detector, thereby reducing the total efficiency to $82\%$. 
The uncertainty on the cross section associated with all efficiencies is 
$< 0.5\%$ \cite{krane}.

Jet energies are corrected \cite{escale}
for the energy response of the D\O\ 
calorimeter to hadrons, the broadening of the hadronic shower, and
energy from multiple interactions, calorimeter noise, and
the underlying event (fragmentation of the spectator partons). 
The response correction increases the $E_{T}$ of jets by 
$22\%$ for measured calorimeter $E_{T}$ of $20\;%
\mathrm{GeV}$, and by $15\%$ for jet $E_{T}$ above 100~GeV. 
The 1\% showering correction recovers the net energy lost when 
hadrons from inside the R=0.7 cone deposit energy outside it 
as they interact within the calorimeter.
Calorimeter noise, from electronics and from uranium
activity, contributes on average $1.6\;\mathrm{GeV}$ 
of $E_{T}$ to each jet. The underlying event contributes 
$0.6\; \mathrm{GeV}$ of $E_{T}$ to each jet at $\sqrt{s}=630\;\mathrm{GeV}$,
compared to $0.9\; \mathrm{GeV}$ at $\sqrt{s}=1800\;\mathrm{GeV}$.
The corrections offset one another, so that a jet's measured $E_{T}$ 
typically increases by 12\% to 14\% after implementing all energy 
scale corrections. Uncertainties in
the corrections for noise and response dominate the systematic uncertainty 
of the final result.

Both detector imperfections and random fluctuations in shower 
development of individual jets within the
calorimeter result in the smearing of a jet's $E_{T}$ about its
true value. The finite $E_{T}$ resolution shifts the observed cross
section to higher $E_{T}$, especially in the most steeply falling regions
of the distribution. The measurement of jet resolution as a function
of $E_{T}$ and the unsmearing procedure follow the steps described
in Ref. \cite{xs_1800}.  The unsmearing correction is larger at 630~GeV than at 
1800~GeV because the cross section is significantly steeper at the
$E_{T}$ values of interest.

\begin{figure}[hbtp]
\epsfxsize=3.25in
\centerline{\epsffile{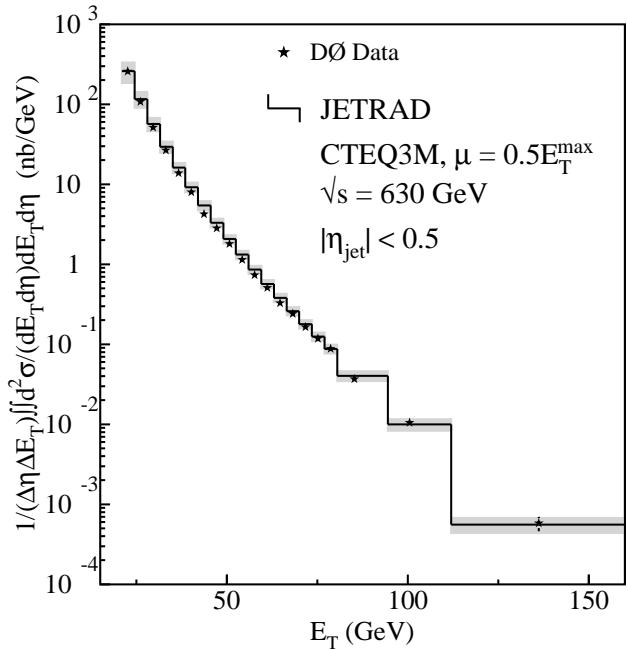}}
 \caption{ The inclusive jet cross section at $\sqrt{s} = 630$~GeV,
integrated over azimuth and averaged over $\left| \eta \right|<0.5$.  The
shaded band corresponds to the systematic uncertainties in the measured
cross section and the solid line shows a prediction from NLO QCD.}
\label{xsec}
\end{figure}

Figure~\ref{xsec} depicts the inclusive jet cross section at 
$\sqrt{s}=630\;\mathrm{GeV}$ 
in the pseudorapidity bin $\left| \eta \right| <0.5$. Each data point 
indicates the $E_{T}$ at which the cross section within that bin has 
its average value. 
The bin widths are chosen to match the bins in $x_{T}$ from the 
$\sqrt{s}=1800\;\mathrm{GeV}$ analysis.   Table~\ref{t_xsec630} reports the bin 
ranges, point positions, and uncertainties. The solid line in Fig.~\ref{xsec}
indicates the result of a calculation using the \textsc{jetrad} next-to-leading-order
(NLO) partonic event generator \cite{jetrad} and the CTEQ3M PDFs \cite{cteq3m}.
The renormalization and factorization scales are set to $\mu =E_{T}^{\rm max}/2$, 
where $E_{T}^{\rm max}$ corresponds to the $E_{T}$ of the leading jet in an event.  

\begin{figure}[hbtp]
\epsfxsize=3.25in
\centerline{\epsffile{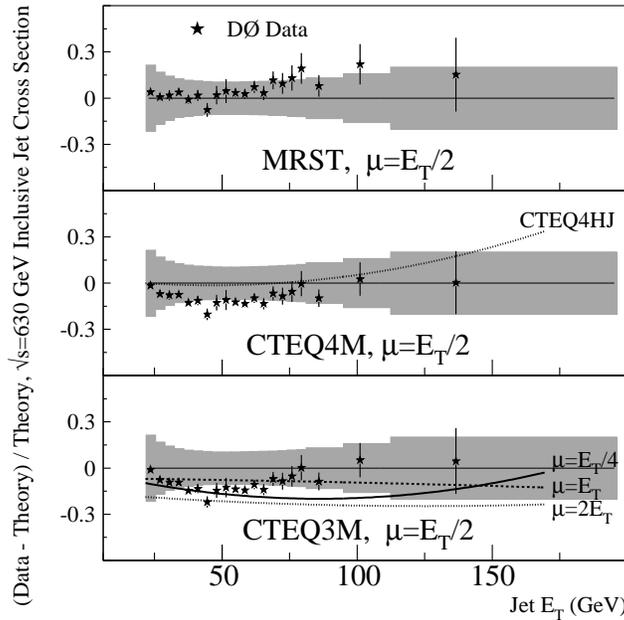}}
 \caption{ The inclusive jet cross section at $\sqrt{s} = 630$~GeV
compared to several NLO QCD predictions. Error bars indicate 
statistical uncertainties and shaded bands correspond to
systematic uncertainties.  The horizontal lines at zero
indicate the baseline prediction that is named in each pane; additional lines indicate 
theoretical variations relative to the baseline.}
\label{dtt630}
\end{figure}

Figure~\ref{dtt630} compares the cross section to the NLO QCD prediction 
in greater detail.  The ``baseline'' renormalization and 
factorization scales are set to $\mu =E_{T}^{\rm max}/2$; additional lines 
in Fig.~\ref{dtt630} indicate the predictions that result from 
changes in either PDF or $\mu$ relative to the baseline
prediction specified for that pane. The shaded regions in Fig.~\ref{dtt630} indicate 
the one standard deviation systematic uncertainty of the measurement,
and the vertical bars indicate the statistical uncertainty. 
The first prediction, generated with the MRST \cite{mrst} 
PDF, is shown to best reproduce the absolute magnitude of the data, but the 
CTEQ4HJ \cite{cteq4mhj} curve in the second pane appears to provide the closest match in shape.
Changing $\mu $ modifies both the normalization and 
the shape of the predictions, as seen in the third pane.  We quantify 
the agreement between the data and the various predictions with a $\chi ^{2}$ 
comparison, as described below.

\label{table1}
\begin{table*}[tbp] \centering
\begin{tabular}
{d@{--}dddcdcd}
\multicolumn{2}{c}{Bin}     & \multicolumn{1}{c}{Plotted}       &   \multicolumn{1}{c}{Plotted} & 
\multicolumn{1}{c}{Cross Sec. (nb/GeV)} & \multicolumn{1}{c}{Cross Sec.} & 
\multicolumn{1}{c}{Ratio} & \multicolumn{1}{c}{Ratio} \\ 
\multicolumn{2}{c}{\Et\ (GeV)}    &  \multicolumn{1}{c}{\Et\ (GeV) }  & \multicolumn{1}{c}{$x_T$}  & 
\multicolumn{1}{c}{$\pm$ Stat. Error} & \multicolumn{1}{c}{ Sys. Error ($\%$)}  &  
\multicolumn{1}{c}{$\pm$ Stat.} & \multicolumn{1}{c}{Sys. ($\%$)} \\
 \hline
   21.0 &   24.5 &   22.6 &   0.072 & $(  2.56 \pm   0.03)\times 10^{ 2}$ &   21.7 & $  1.72\pm  0.03 $ &   12.7\\
   24.5 &   28.0 &   26.1 &   0.083 & $(  1.07 \pm   0.02)\times 10^{ 2}$ &   17.2 & $  1.64\pm  0.04 $ &    9.7\\
   28.0 &   31.5 &   29.6 &   0.094 & $(  5.14 \pm   0.16)\times 10^{ 1}$ &   14.6 & $  1.62\pm  0.06 $ &    8.0\\
   31.5 &   35.0 &   33.1 &   0.105 & $(  2.67 \pm   0.05)\times 10^{ 1}$ &   13.0 & $  1.67\pm  0.03 $ &    7.0\\
   35.0 &   38.5 &   36.7 &   0.116 & $(  1.37 \pm   0.04)\times 10^{ 1}$ &   12.1 & $  1.57\pm  0.04 $ &    6.3\\
   38.5 &   42.0 &   40.2 &   0.127 & $(  7.96 \pm   0.27)\times 10^{ 0}$ &   11.5 & $  1.59\pm  0.06 $ &    6.0\\
   42.0 &   45.5 &   43.7 &   0.139 & $(  4.24 \pm   0.20)\times 10^{ 0}$ &   11.2 & $  1.48\pm  0.07 $ &    5.8\\
   45.5 &   49.0 &   47.2 &   0.150 & $(  2.83 \pm   0.16)\times 10^{ 0}$ &   11.0 & $  1.63\pm  0.09 $ &    5.5\\
   49.0 &   52.5 &   50.7 &   0.161 & $(  1.81 \pm   0.13)\times 10^{ 0}$ &   10.9 & $  1.64\pm  0.12 $ &    5.4\\
   52.5 &   56.0 &   54.2 &   0.172 & $(  1.14 \pm   0.03)\times 10^{ 0}$ &   10.9 & $  1.64\pm  0.04 $ &    5.4\\
   56.0 &   59.5 &   57.7 &   0.183 & $(  7.35 \pm   0.21)\times 10^{-1}$ &   11.0 & $  1.62\pm  0.05 $ &    5.4\\
   59.5 &   63.0 &   61.2 &   0.194 & $(  5.07 \pm   0.17)\times 10^{-1}$ &   11.1 & $  1.67\pm  0.06 $ &    5.4\\
   63.0 &   66.5 &   64.7 &   0.205 & $(  3.29 \pm   0.14)\times 10^{-1}$ &   11.3 & $  1.60\pm  0.07 $ &    5.5\\
   66.5 &   70.0 &   68.2 &   0.216 & $(  2.42 \pm   0.12)\times 10^{-1}$ &   11.5 & $  1.74\pm  0.09 $ &    5.5\\
   70.0 &   73.5 &   71.7 &   0.228 & $(  1.64 \pm   0.10)\times 10^{-1}$ &   11.8 & $  1.69\pm  0.10 $ &    5.6\\
   73.5 &   77.0 &   75.2 &   0.239 & $(  1.18 \pm   0.08)\times 10^{-1}$ &   12.1 & $  1.78\pm  0.13 $ &    5.8\\
   77.0 &   80.5 &   78.7 &   0.250 & $(  8.79 \pm   0.72)\times 10^{-2}$ &   12.4 & $  1.81\pm  0.15 $ &    5.9\\
   80.5 &   94.5 &   85.2 &   0.271 & $(  3.69 \pm   0.23)\times 10^{-2}$ &   13.6 & $  1.74\pm  0.11 $ &    6.4\\
   94.5 &  112.0 &  100.5 &   0.319 & $(  1.05 \pm   0.11)\times 10^{-2}$ &   16.2 & $  1.85\pm  0.20 $ &    7.7\\
  112.0 &  196.0 &  136.2 &   0.432 & $(  5.81 \pm   1.19)\times 10^{-4}$ &   20.4 & $  1.83\pm  0.38 $ &    9.7\\
\end{tabular}
\caption{Inclusive jet cross section at
$\sqrt{s}=630~\mathrm{GeV}$ and the ratio of dimensionless cross sections
$f^{630}(x_{T})/f^{1800}(x_{T})$, where 
$f\left( x_{T}\right) =%
E_{T}^{4}\cdot E\frac{d^{3}\sigma }{d^{3}p}$ and $x_{T}=2E_{T}/\sqrt{s}$.  The cross 
sections are all integrated over azimuth and averaged in the range $\left| \eta \right| <0.5$.
\label{t_xsec630}}%
\end{table*}%

\begin{figure}[hbtp]
\epsfxsize=3.25in
\centerline{\epsffile{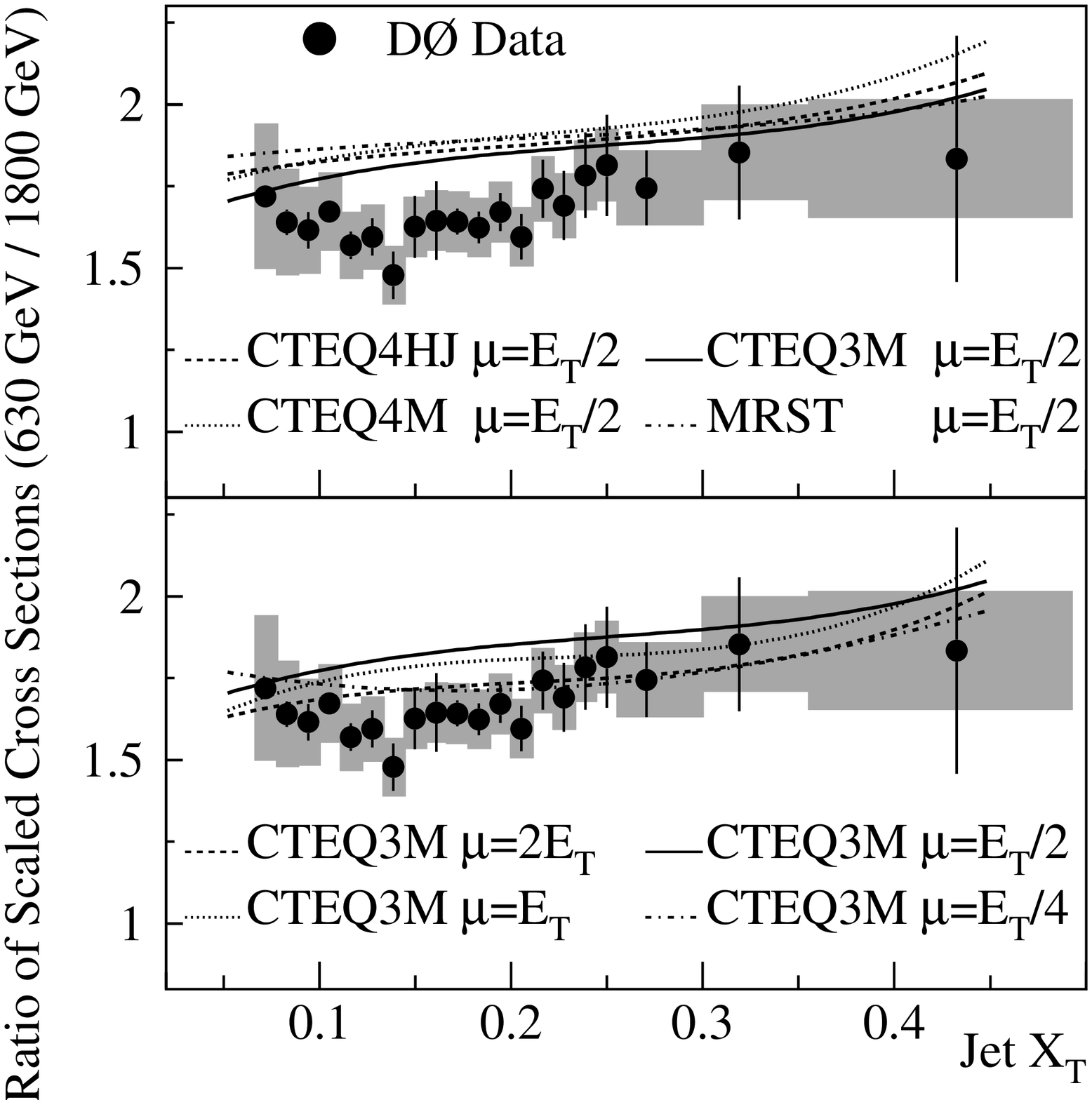}}
 \caption{Ratio of dimensionless jet cross sections (numerator
$\sqrt{s} = 630$~GeV, denominator $\sqrt{s} = 1800$~GeV) compared to 
NLO QCD as given by \textsc{jetrad}.  Error bars indicate 
statistical uncertainties and shaded bands correspond to
systematic uncertainties.  }
\label{ratio}
\end{figure}

Combining the results from this Letter with those of
Ref. \cite{xs_1800}, Fig.~\ref{ratio} displays the ratio of
dimensionless jet cross sections as a function
of $x_{T}$.  The observed ratio ranges from 1.48 
to 1.85, depending on the value of $x_{T}$.  The largest uncertainties 
arise from the corrections for response and noise, and the rest 
primarily from resolution and luminosity.
Although the systematic errors on the individual measurements range from
10\% to as much as 30\%, strong correlations reduce the
uncertainty on the ratio to values as small as $\pm 5.4\%$. The two final
columns of Table~\ref{t_xsec630} provide the numerical results
for the ratio.

As shown in Fig.~\ref{ratio}, NLO QCD predictions for the ratio lie
systematically above the data throughout most of the measured $x_{T}$ range,
in particular between $x_{T}$ of 0.1 and 0.2, where the ratio has the smallest
statistical uncertainty. Choice of PDF has little effect on the prediction --- only the
renormalization/factorization scales change the prediction appreciably. 

A covariance matrix $\chi ^{2}$ comparing data and theory provides a measure 
of the probability that the theory describes the observed results.
To verify that our covariance matrix, built mostly from correlated
systematic uncertainties, produces results that are consistent with
a standard $\chi ^{2}$ distribution with 20 degrees of freedom, we 
generated an ensemble of 20 million experiments using a Monte Carlo program.  
Each statistical and systematic error was simulated and varied 
randomly using appropriate correlations in $E_{T}$ and $\sqrt{s}$.  
Systematic errors were not necessarily assumed to be Gaussian distributed; 
some numbers were drawn from uniform probability distributions,
as appropriate. The $\chi ^{2}$ comparisons (between the 
original input and each of the final, randomly-varied distributions) 
is in excellent agreement with the shape of the 
$\chi ^{2}$ function for 20 degrees of freedom.
We find that the standard probability obtained from an
integral of the $\chi ^{2}$ distribution provides an appropriate 
vehicle for comparing data with predictions.

Table \ref{t_results} reports both the $\chi ^{2}$ values and 
the $\chi ^{2}$ probabilities for the comparison of the data with different
NLO QCD predictions.  The inclusive jet cross section at $\sqrt{s}=630$~GeV
is consistent with all the tested PDFs and $\mu$ scales, with but
two exceptions.  For the ratio of cross sections, there is no 
significant difference in shape between data and theory, and 
essentially all predictions lie within an acceptable range. 
The overall results in Table \ref{t_results} indicate reasonable 
agreement between the ratio and NLO QCD.

We performed a second test to quantify the observed difference in the
absolute magnitudes of the predicted and observed ratios, without 
particular regard to the 
shapes of the distributions.  Using the covariance matrix and assuming 
that the value of the ratio is a constant with respect to $x_{T}$, we
found the best-fit horizontal line for the data.  The $\chi ^{2}$ value 
that results from a comparison of this single point to the 
equivalently-calculated theory point yields the probabilities listed in the 
final column of Table \ref{t_results}. In every case, discarding the 
information on shape in favor of a comparison of absolute magnitude 
results in poorer agreement between data and theory, particularly
for the often-favored scale of $\mu = E_{T}^{\rm max}/2$.

In conclusion, we have measured the inclusive jet 
cross section at two center-of-mass energies, 630~GeV and 1800~GeV.  
Both the published data at 1800~GeV \cite{xs_1800} and the data
presented here at 630~GeV are generally well-described by NLO QCD,
with the exception of predictions using CTEQ3M($2E_{T}^{\rm max}$) and MRSTGD PDFs.
In the ratio of dimensionless cross sections at the two energies, 
experimental uncertainties are much smaller and differences in the predictions
from choice of PDF are less important.  NLO predictions for the ratio exhibit satisfactory 
agreement with the shape of the observed ratio.  In terms of only the magnitude
however, the absolute values of the predictions lie significantly 
higher than the data, especially for the standard scale $\mu =E_{T}^{\rm max}/2$.

\begin{table}[tbp] \centering%
\begin{tabular}{cccccccc}
PDF & $\mu$ & 
\multicolumn{2}{c}{630~GeV C. Sec.}  & \multicolumn{2}{c}{Ratio} & \multicolumn{2}{c}{Norm.} \\
 &  &  $\chi ^{2}$ & Prob. & $\chi ^{2}$ & Prob. & $\chi ^{2}$ & Prob. \\ \hline
 & 2$\cdot E_{T}^{\rm max}$   & 40.5 & 0.43\% & 17.9 & 59.4\% & 3.33 & 6.81\% \\ 
 & $E_{T}$          & 25.9 & 16.8\% & 21.6 & 36.2\% & 7.13 & 0.76\% \\ 
CTEQ3M & $E_{T}^{\rm max}/2$  & 30.4 & 6.37\% & 20.5 & 42.5\% & 9.56 & 0.20\% \\ 
 & $E_{T}^{\rm max}/4$        & 27.5 & 12.2\% & 15.1 & 77.4\% & 1.45 & 22.93\% \\ \hline
CTEQ4M & $E_{T}^{\rm max}/2$  & 24.1 & 23.8\% & 22.4 & 31.9\% & 10.67 & 0.11\% \\ 
CTEQ4HJ & $E_{T}^{\rm max}/2$ & 18.9 & 52.5\% & 21.0 & 40.0\% & 13.21 & 0.03\% \\ \hline
MRST    & $E_{T}^{\rm max}/2$ & 22.6 & 30.7\% & 22.2 & 33.0\% & 12.60 & 0.04\% \\
MRSTGU  & $E_{T}^{\rm max}/2$ & 14.9 & 78.2\% & 19.5 & 48.7\% & 11.07 & 0.09\% \\
MRSTGD  & $E_{T}^{\rm max}/2$ & 51.8 & $0.012\%$ & 24.1 & 23.9\% & 12.92 & 0.03\% \\
\end{tabular}
\caption{$\chi^{2}$ comparisons for the cross section at $\sqrt{s}=630$~GeV (20
degrees of freedom), the ratio of cross sections (20 degrees of freedom), and a 
comparison for the ratio involving only the absolute magnitude (one degree of 
freedom).\label{t_results}}%
\end{table}

%
We thank the staffs at Fermilab and at collaborating institutions 
for contributions to this work, and acknowledge support from the 
Department of Energy and National Science Foundation (USA),  
Commissariat  \` a L'Energie Atomique and
CNRS/Institut National de Physique Nucl\'eaire et 
de Physique des Particules (France), 
Ministry for Science and Technology and Ministry for Atomic 
   Energy (Russia),
CAPES and CNPq (Brazil),
Departments of Atomic Energy and Science and Education (India),
Colciencias (Colombia),
CONACyT (Mexico),
Ministry of Education and KOSEF (Korea),
CONICET and UBACyT (Argentina),
A.P. Sloan Foundation,
and the A. von Humboldt Foundation.

\end{document}